\documentclass[twocolumn,showpacs,preprintnumbers,amsmath,amssymb,superscriptaddress,groupedaddress,prb]{revtex4}

\usepackage{epsfig}                                                            
\usepackage{graphicx}
\usepackage{dcolumn}
\usepackage{color}
\usepackage{bm}
\usepackage{subfigure} 
\usepackage{epsfig} 

\begin{document}

\title{\bf Origin of the asymmetric exchange bias in BiFeO$_3$-Bi$_2$Fe$_4$O$_9$ nanocomposite}

\author {Tuhin Maity} \affiliation {Micropower-Nanomagnetics Group, Microsystems Center, Tyndall National Institute, University College Cork, Lee Maltings, Dyke Parade, Cork, Ireland}
\author {Sudipta Goswami} \affiliation {Nanostructured Materials Division, CSIR-Central Glass and Ceramic Research Institute, Kolkata 700032, India}
\author {Dipten Bhattacharya} 
\email{dipten@cgcri.res.in} \affiliation {Nanostructured Materials Division, CSIR-Central Glass and Ceramic Research Institute, Kolkata 700032, India}
\author {Saibal Roy}
\email{saibal.roy@tyndall.ie} \affiliation {Micropower-Nanomagnetics Group, Microsystems Center, Tyndall National Institute, University College Cork, Lee Maltings, Dyke Parade, Cork, Ireland} 

\date{\today}

\begin{abstract}
We show from detailed magnetometry across 2-300 K that the BiFeO$_3$-Bi$_2$Fe$_4$O$_9$ nanocomposite offers a unique spin morphology where superspin glass (SSG) and dilute antiferromagnet in a field (DAFF) coexist at the interface between ferromagnetic Bi$_2$Fe$_4$O$_9$ and antiferromagnetic BiFeO$_3$. The coexisting SSG and DAFF combine to form a local spin texture which gives rise to a path-dependent exchange bias below the spin freezing temperature ($\sim$29K). The exchange bias varies depending on the protocol or path followed in tracing the hysteresis loop. The exchange bias has been observed below blocking temperature (T$_B$$\sim$60K) of the superparamagnetic Bi$_2$Fe$_4$O$_9$. The conventional exchange bias (CEB) increases nonmonotonically as temperature decreases. The magnitude of both exchange bias (H$_E$) and coercivity (H$_C$) increase with decrease in temperature and are found to be asymmetric below 20K depending on the path followed in tracing the hysteresis loop and bias field. The local spin texture at the interface between ferromagnetic and antiferromagnetic particles generates a nonswitchable unidirectional anisotropy along the negative direction of the applied field. The influence of this texture also shows up in "asymmetric" jumps in the hysteresis loop at 2 K which smears off at higher temperature. The role of the interface spin texture in yielding the path dependency of exchange bias is thus clearly delineated.  
\end{abstract} 
\maketitle

In a period spanning more than five decades now, exchange bias effect has been observed in a multitude of magnetic heterostructures: (i) ferromagnet-antiferromagnet;\cite{Bean} (ii) ferromagnet-spin glass;\cite{Ali} (iii) ferrimagnet-antiferromagnet;\cite{Wolf} (iv) ferrimagnet-ferromagnet;\cite{Cain} (v) ferromagnet-spin glass-antiferromagnet\cite{Maity} etc. Conventionally, the exchange bias - measured by shift of the magnetic hysteresis loop along the field axis - requires pre-biasing of the interface moment via a field cooling protocol from above the magnetic transition point.\cite{Schuller} This pre-biasing sets the unidirectional anisotropy by breaking the symmetry of the interface moment. Exchange bias has also been observed, spontaneously, when even in the absence of pre-biasing, unidirectional anisotropy sets in under the first field of loop evaluation,\cite{Saha},\cite{Maity1} where the sample is cooled down from above the transition point under zero field. In the present work, a remarkable effect emerges when a unidirectional anisotropy along a particular direction with respect to the direction of the applied field is set due to the field cooling under so-called conventional exchange bias. The unidirectional anisotropy thus set under field turns out to be in negative direction  which gives rise to asymmetric exchange bias. This nonswitchable unidirectional anisotropy seems to lie at the heart of the observed path or protocol dependency of exchange bias. What kind of magnetic heterostructure can generate such an asymmetric exchange bias? We show, in this paper, from detailed magnetometry that the path dependency is observed in a unique spin morphology where superspin glass (SSG) and dilute antiferromagnet in a field (DAFF) structures coexist at the interface between finer and ferromagnetic Bi$_2$Fe$_4$O$_9$ and coarser and antiferromagnetic BiFeO$_3$ in a nanocomposite of BiFeO$_3$-Bi$_2$Fe$_4$O$_9$. They appear to form a local spin texture at the interface which yields the asymmetry for the onset of universal unidirectional anisotropy at the interface along a particular direction. This local spin texture is also instrumental in yielding sharp jumps - rather asymmetrically - in the hysteresis loop measured below 20 K due to the existence of FM-SSG-DAFF-AFM bias coupling at the nanoscale. We, therefore, offer here a direct correlation between the jumps in the hysteresis loops at low temperature and the  symmetry breaking of the interface moment which sets nonswitchable universal unidirectional anisotropy and turns out to be giving rise to the observed path dependency of the exchange bias.     

\begin{figure*}[!htp]
  \begin{center}
    \subfigure[]{\includegraphics[scale=0.45]{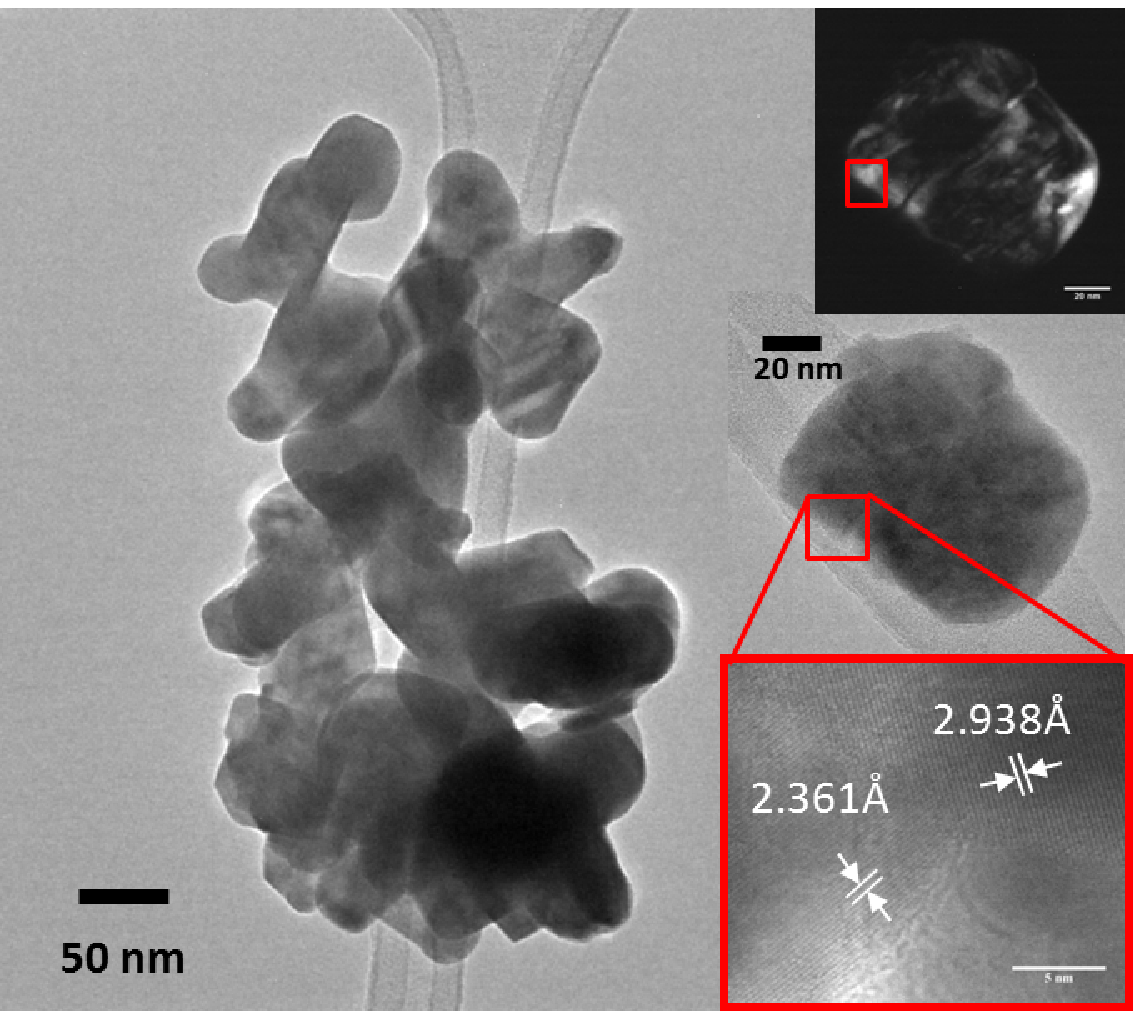}}
    \subfigure[]{\includegraphics[scale=0.75]{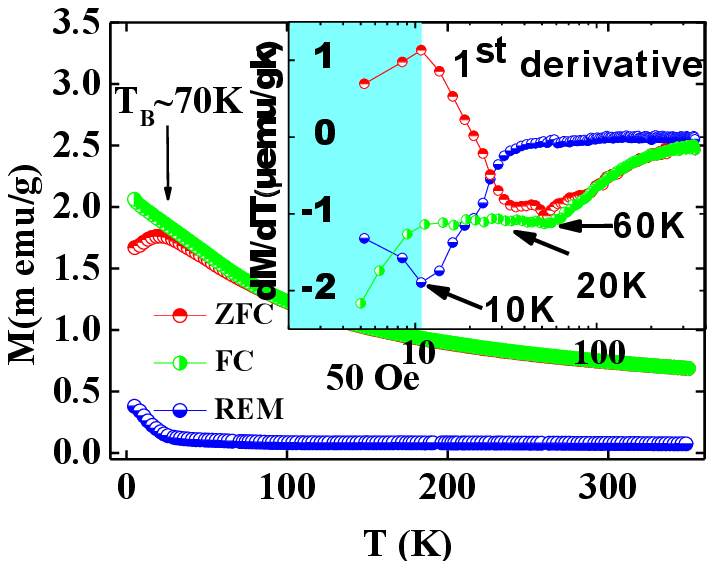}}  
	 \subfigure[]{\includegraphics[scale=0.80]{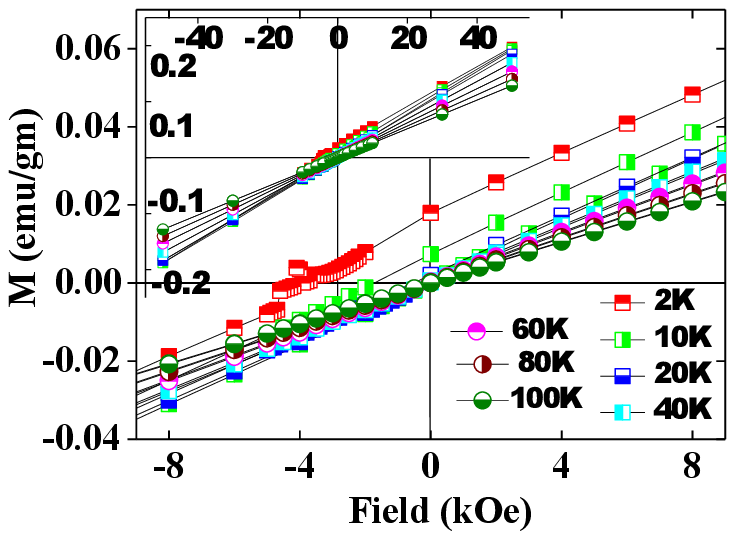}}
    \subfigure[]{\includegraphics[scale=0.20]{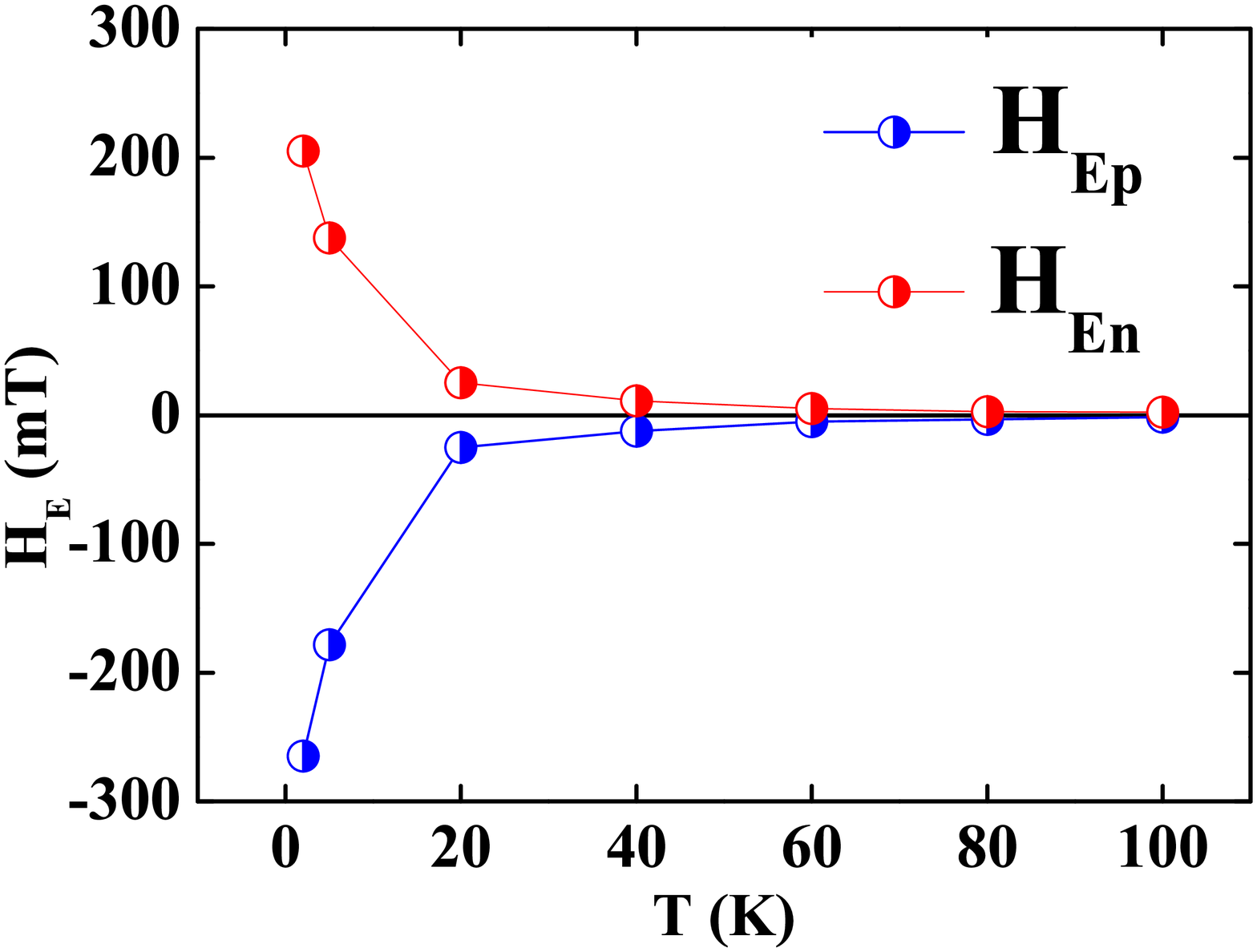}} 
	 \subfigure[]{\includegraphics[scale=0.20]{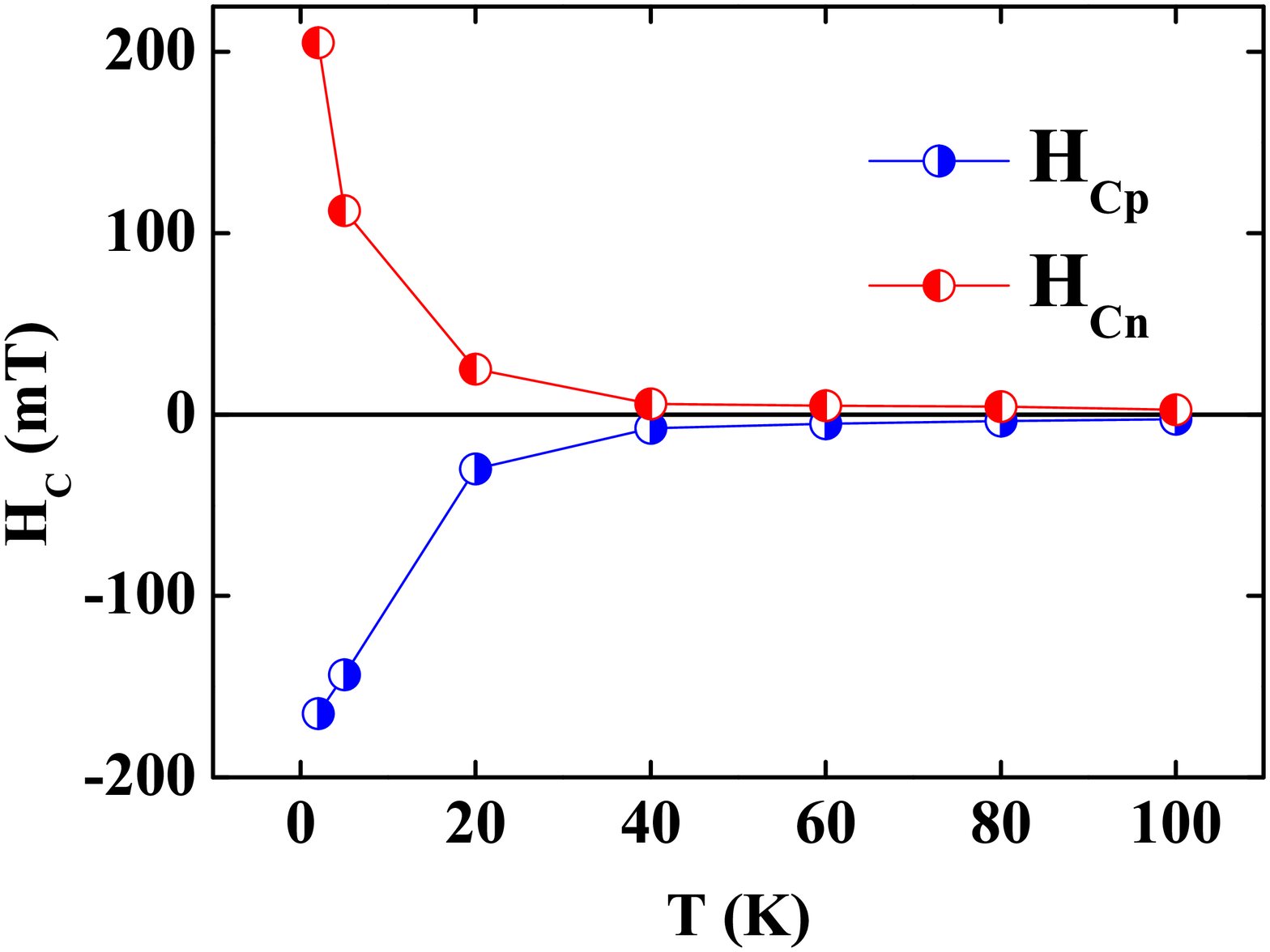}}
    \subfigure[]{\includegraphics[scale=0.50]{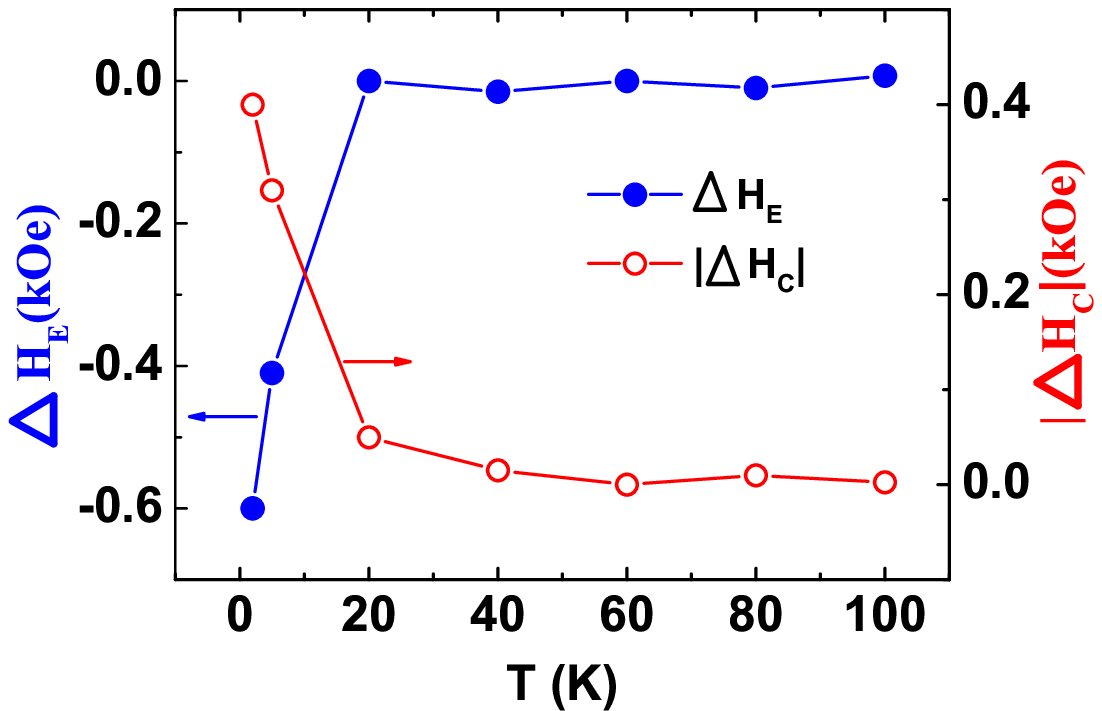}} 
  \end{center}
  \caption{(color online) (a) A representative bright field TEM image of the nanocomposite; top inset shows a representative bright-field TEM image of an interface; second inset shows the dark-field image of the region while the bottom inset shows the HRTEM image of different orientation of the lattice fringes at the interface. (b) The zero-field cooled, field-cooled, and remanent magnetization versus temperature plots; inset shows the $\frac{dM}{dT}$ versus T plots; (c) the hysteresis loop shifts at different temperatures showing the exchange bias; portion near the origin is blown up; inset shows the full loops; the temperature dependence of (d) exchange bias and (e) coercivity and the asymmetry of (f) exchange bias and  coercivity across 5-100 K.}
\end{figure*}

The experiments have been carried out on multiple nanocomposite samples of BiFeO$_3$-Bi$_2$Fe$_4$O$_9$. The nanocomposite was synthesized  by sonochemical route \cite{Goswami} and characterized by high resolution powder x-ray diffraction, transmission and high resolution transmission electron microscopy, selected area electron diffraction methord etc. The sample is found to contain $\sim$90 vol\% BiFeO$_3$ and $\sim$10 vol\% Bi$_2$Fe$_4$O$_9$. The average particle sizes are $\sim$57 nm for BiFeO$_3$ and $\sim$13 nm for Bi$_2$Fe$_4$O$_9$. The magnetic measurements have been carried out in a SQUID magnetometer (MPMS XL 5, Quantum Design) across a temperature range 2-350 K and under a maximum applied field 50 kOe. Prior to each measurement both the sample and the superconducting magnet were demagnetized following an appropriate protocol. In Fig. 1, the major results from the magnetic measurements including HRTEM images are shown. In Fig.1a nanocomposite of BFO is shown. Both the phases BiFeO$_3$ and Bi$_2$Fe$_4$O$_9$ are identified by dark field, bright field imaging and corresponding interfaces (Inset figures). Fig.1b shows the zero-field cooled and field cooled (ZFC and FC) magnetization ($M$) versus temperature ($T$) plots. Inset shows the $\frac{dM}{dt}$ versus $T$ plots which help in identifying the blocking temperature $T_B$. The change in slope of the $\frac{dM}{dt}$ versus $T$ plots below 10 K signifies weak ferromagnetism.\cite{Rao, Zhang} Fig. 1c shows the hysteresis loops measured across 2-100 K. The portion near the origin is blown up to show the asymmetric shift of the loops along the field axis. The exchange bias ($H_E$) observed in this case is conventional as the measurement has been carried out following field cooling under +50 kOe and -50 kOe. Importantly, the extent of exchange bias turns out to be dependent on the sign of the field applied during field cooling and also on the path followed in tracing the loop: +50 kOe $\rightarrow$ 0 $\rightarrow$ -50 kOe $\rightarrow$ 0 $\rightarrow$ +50 kOe (positive) or -50 kOe $\rightarrow$ 0 $\rightarrow$ +50 kOe $\rightarrow$ 0 $\rightarrow$ -50 kOe (negative). In Fig. 1d, we show the asymmetric or path-dependent exchange bias ($H_E$: $H_Ep$-Positive, $H_En$-Negative) and coercivity ($H_C$: $H_Cp$-Positive, $H_Cn$-Negative) as a function of temperature. The exchange bias is given by $H_E$ = ($H_{c1}$+$H_{c2}$)/2 and the coercivity is given by $H_C$ = ($H_{c1}$-$H_{c2}$)/2 where $H_{c1}$ and $H_{c2}$ are the fields at which the magnetization reaches zero during the tracing of forward and reverse branches of the hysteresis loop. Further ananlysis of the hysteresis loops also reveals a vertical shift along the magnetization axis. The vertical shift was earlier observed \cite{Nogues} to be associated with the exchange bias and was resulting from induced net moment. While ferromagnetic coupling across the interface yields a positive shift, antiferromagnetic coupling results in a negative shift. Consistent with the earlier observation \cite{Nogues}, positive shift here is associated with negative exchange bias. However the asymmetry in the vertical shift was not observed between the loops traced via positive and negative paths at 2K after cooling with +/- 5T bias field. Finally the temperature dependences of the net exchange bias ($\Delta$H$_E$) and coercivity ($\Delta$H$_C$) are shown in the Figs. 1e and 1f, respectively. Interestingly, both the $H_E$ and $\Delta$H$_E$ (and likewise $H_C$ and $\Delta$H$_C$) increase sharply with the decrease in temperature below 20 K. The $T_B$, however, is 60 K. In between 20 and 60 K, exchange bias, coercivity, and their asymmetry are small and exhibit a rather weak temperature dependence. In order to probe this observation further, we have carried out ac susceptibility measurements as well. In Fig. 2, we show the complex ac susceptibility versus temperature plots for different frequencies. It appears that a distinct spin freezing transition takes place around 20 K.  The peak temperatureT$_f$ ($w$)  shifts towards higher temperature and linearly increases with log $w$ with the increase of frequencies (inset Fig.2)  which is a clear signature of spin glass behavior.\cite{Mydosh} The frequency sensitivity K of T$_f$ ($w$) has been calculated to be 4.7. This frequency dependence of T$_f$ ($w$) is described as conventional “slowing down” of spin dynamics which results the irreversibility in the spin glass.\cite{De} The frequency dependence (inset of Fig. 2) of the peak temperature follows Vogel-Fulcher pattern (Vogel-Fulcher freezing temperature was earlier found\cite{Singh} to be $\sim$29.4 K for BiFeO$_3$). Previous reports suggest that at low temperature BFO possess  low temperature spin glass ordering which leads to the increase of H$_B$ and H$_C$ at low temperature.\cite{Singh} This result shows that the onset of spin-freezing transition has a strong bearing upon the exchange bias and its asymmetry. 

\begin{figure}[!h]
  \begin{center}
    \includegraphics[scale=0.30]{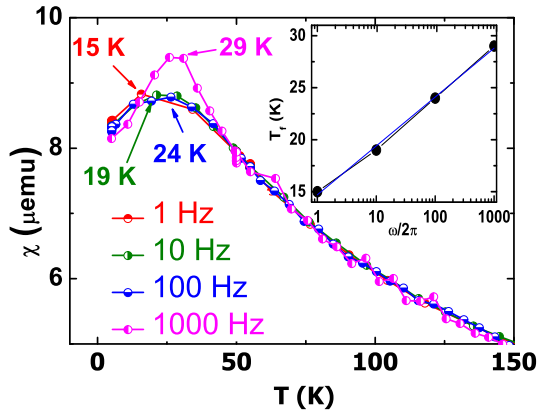} 
    \end{center}
  \caption{(color online) The complex ac susceptibility versus temperature plot at different frequencies; inset shows the shift in the peak temperature with the frequency. }
\end{figure}

In order to probe the spin morphology of the system, we have carried out detailed magnetometry across 2-300 K. We have studied the relaxation of the moment at 2 K over a time span of 3600s under both +50 kOe and -50 kOe. The sample was first cooled down from 350 K to 2 K under zero field and then +50 kOe was applied and the time dependence of the moment was measured for 3600s. After that the field was ramped down to -50 kOe and the magnetization was measured for 3600s. Again the field was ramped back to +50 kOe and the measurement was repeated for 3600s. The results of these three measurements are plotted in Fig. 3a. They clearly show an upward creep signifying incoherent rotation of the ferromagnetic moment because of the presence of superspin glass (SSG) at the interface.\cite{Sasaki,Kodama} Additionally, in this high field relaxation process we observed that the amount of variation in magnitude of moments (M) in 3600 sec relaxation time is almost same for alternative fields. It indicates the uniaxiality (UA) of ferromagnetic grains. Thus the asymmetry in exchange bias does not come from the FM part of the composite. The possible reason of  asymmetry could be the existence of random anisotropy  at the interface of BiFeO3- Bi2Fe4O9 and interactions with uniaxial anisotropy of very small size of the ferromagnetic domains through a non-trival interface spin structure where the exchange bias coupling freezes below Vogel–Fulcher freezing temperature. The presence of superspin glass was also investigated in memory effect on ZFC magnetization measured by well-designed stop-and-wait protocol. The characteristic peak in the differential moment versus temperature plot at a temperature at which the measurement was stopped and waited for 10$^4$s signifies presence of SSG in the system. We further carried out isothermal remanence and thermoremanence measurements at 5 K. For the thermoremanence measurement the sample was cooled down from room temperature to 5 K under a specific field and then the field was removed. The remanent moment was measured immediately. The isothermal remanence was measured following zero field cooling. In this case, the sample was brought down to 5 K from room temperature under zero field and momentarily a field was applied at 5 K. Then the field was removed and the magnetization was measured. The field dependence of both the thermo and isothermal remanence at 5 K is shown in Fig. 3b. Very interestingly, the patterns follow closely those expected for a two-dimensional dilute antiferromagnet in a field (DAFF).\cite{Benitez} While the isothermal remanence exhibits a weak field dependence, the thermoremanence follows $\propto$ H$^{\nu_H}$ pattern where $\nu_H$ = 0.64. In the case of spin glass,\cite{Tholence} the isothermal remanence curve increases with field relatively sharply and exhibits a peak at an intermediate field and meets the thermoremanence curve and then both saturate at higher field. For superparamagnetic system,\cite{Manna} the thermoremanence curve increases with field quite rapidly. For the present case, the results of thermo and isothermal remanence measurements indicate presence of two-dimensional DAFF. The high field relaxation process and memory effect on ZFC magnetization, on the other hand, signify simultaneus presence of SSG. The overall spin morphology, therefore, appears to be consisting of four components: ferromagnetic (FM) and antiferromagnetic (AFM) cores and interfaces vitiated by SSG and DAFF shells.

\begin{figure}[!ht]
  \begin{center}
   \subfigure[]{\includegraphics[scale=0.25]{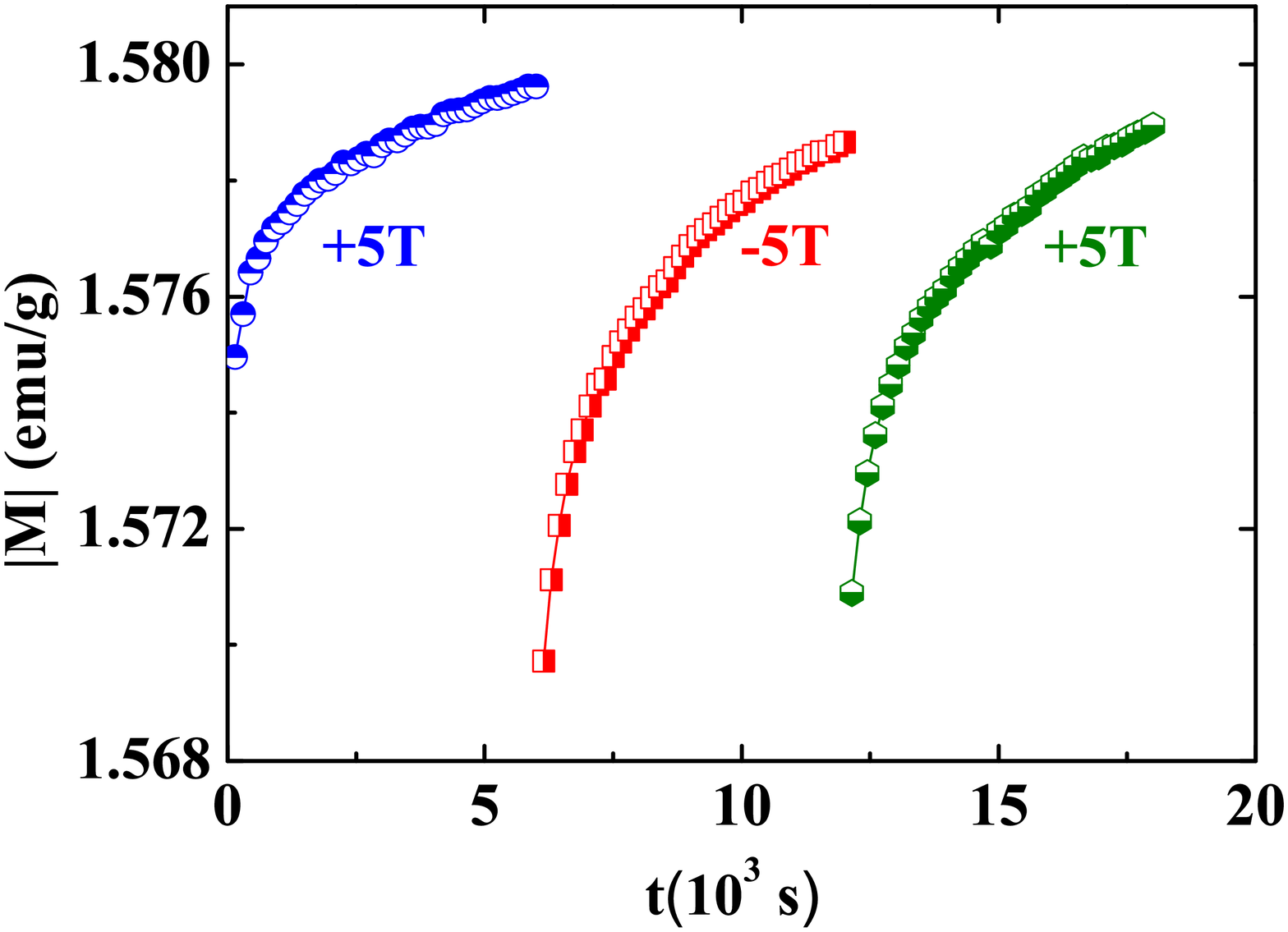}}
    \subfigure[]{\includegraphics[scale=0.25]{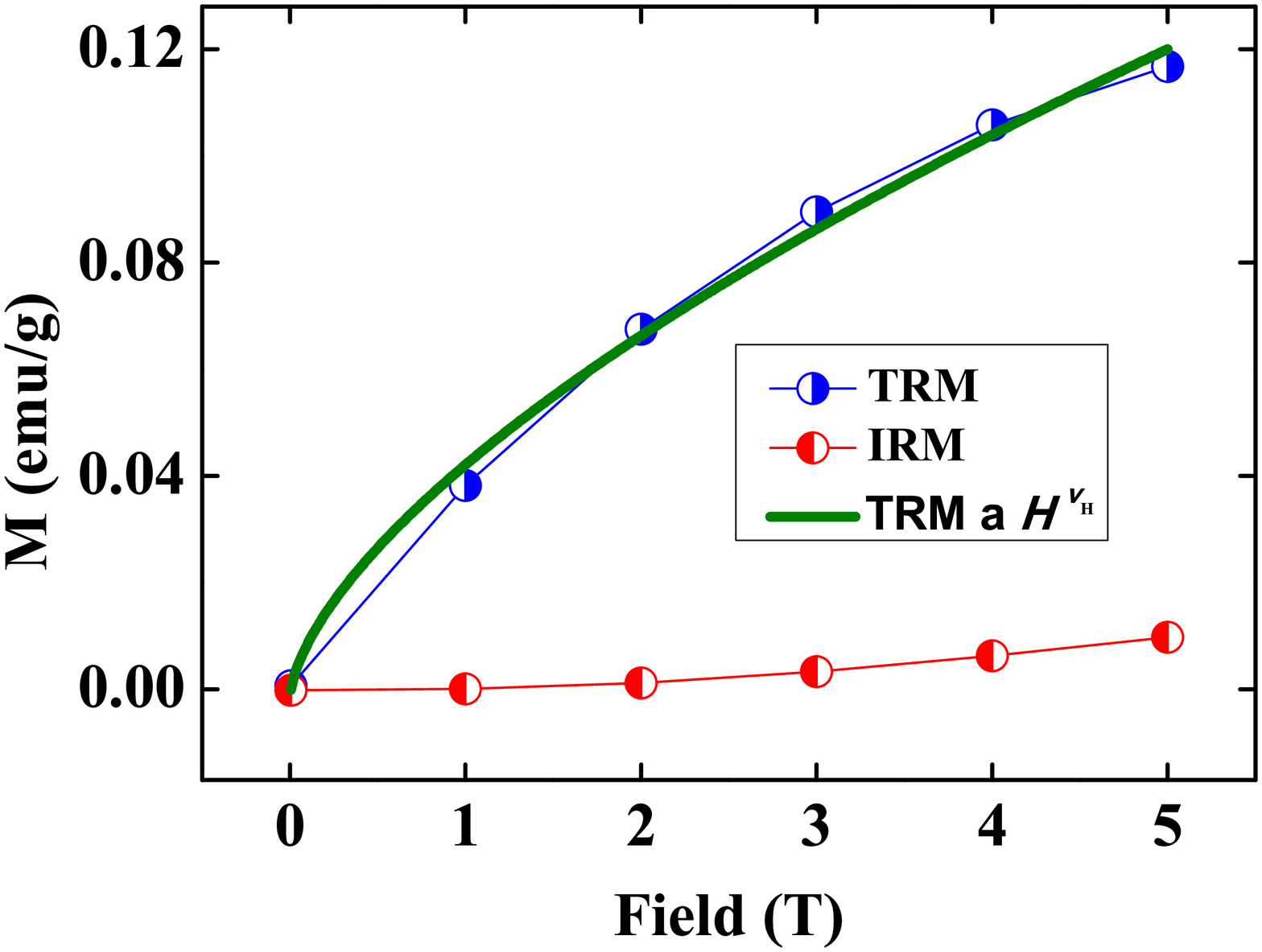}}  
    \end{center}
  \caption{(color online) (a) The relaxation of the magnetization measured alternatively under +50 and -50 kOe at 5 K; (b) field dependence of the thermo and isothermal remanence at 5 K. }
\end{figure}

Finally, we have measured the hysteresis loops with higher resolution at 2 K starting from +50 kOe and from -50 kOe. The loops, especially the portion near the origin, are shown in Fig. 4. Quite conspicuous are the sharp jumps in the loops. Such jumps have earlier been observed in systems containing inhomogeneities and thus random anisotropy.\cite{Calderon, Marcano-1, Marcano-2} In fact, both by experimental and theoretical work, it has been shown that depending on the strength of the random anisotropy with respect to the exchange coupling, several jumps might be seen in hysteresis loops at low temperature. Because of thermal perturbation, they smear off at higher temperature. In the present case too, loops measured at higher temperature (5 K) exhibit lesser number of jumps and complete smearing off eventually at temeprature higher than that. However, there is an interesting distinction between the jumps observed in a ferromagnetic system containing purely random anisotropy because of inhomogeneities and the jumps observed here. The number of jumps observed, in the present case, are different in two different branches of a particular loop. The inversion symmetry, normally observed in systems containing purely random anisotropy,\cite{Marcano-1} is broken here. While lesser jumps (J$_1$, J$_2$) could be seen in the forward branch, more jumps (J'$_1$, J'$_2$, J'$_3$)  are conspicuous in the reverse branch (Fig. 4). This is true for both the loops - whether the loop has been traced starting from +50 kOe or -50 kOe. There is, however, conspicuous inversion symmetry in the jump structure in between two forward and two reverse branches of the loops traced strating from +50 kOe and -50 kOe. It is possible to notice that the branch on the extreme left (blue line) is a mirror image of the branch on the extreme right (red line) of Fig. 4. Likewise, the inner blue and red branches are also mirror images of each other. Such an asymmetric pattern of jumps for a particular loop yet a symmetric one between the loops traced via two different protocols is remarkable and has not been reported earlier. This observation clearly points out that there could be a correlation between this symmetry of the jump pattern and the protocol dependency of the exchange bias. It has been argued below that this results from a local topological spin texture at the interface which, in turn, gives rise to a strong universal unidirectional anisotropy of the interface moment along negative field direction.

\begin{figure}[!h]
  \begin{center}
    \includegraphics[scale=0.70]{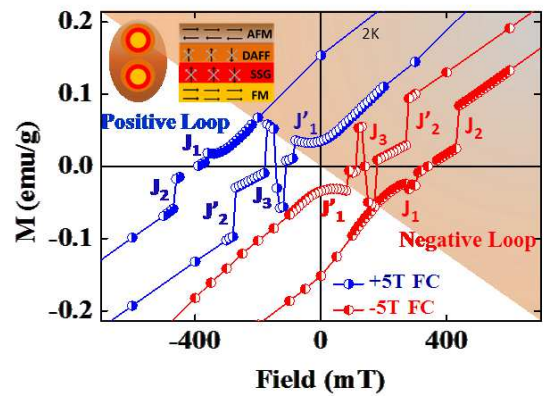} 
    \end{center}
  \caption{(color online) The hysteresis loops measured at 2 K following different protocols starting with +50 kOe (blue line) and -50 kOe (red line); anomalous jump structures could be seen in different branches of the loops; inset shows the spin structure. }
\end{figure}

It has been shown that random anisotropy due to inhomogeneities leads to jumps in the hysteresis loop at a very low temperature (100-500 mK) as a result of avalanches in domain flipping.\cite{Calderon,Marcano-2} The random anisotropy field competes with the applied field and as and when the spins are aligned with the local anisotropy, the avalanche takes place. The role of local anisotropy has been addressed both by site-centric local random field model\cite{Calderon} as well as cluster model.\cite{Marcano-2} The hamiltonian considers the exchange coupling among the spins, local random field due to anisotropy, and the applied field. In the case of the present system, exhibiting exchange bias, exchange coupling interaction across the interface between the BiFeO$_3$ and Bi$_2$Fe$_4$O$_9$ particles should also be considered. This is strongly influenced by the SSG and DAFF layers at the interface. While BiFeO$_3$ crystallizes in rhombohedral structure with R3c space group, Bi$_2$Fe$_4$O$_9$ crystallizes in orthorhombic structure with Pbam space group. The interface, therefore, creates a certain topology which induces, at least, a local spin texture even though globally the salient features of the SSG and DAFF are retained. Therefore, in presence of such local spin texture both random and textured anisotropy field compete with the applied field and the exchange coupling interaction across the interface. The domain flipping and avalanche along two different pathways is not identically influenced. The textured anisotropy creates a self generating interfatial moment which acts upon the exchange coupling interaction in between BiFeO$_3$ and Bi$_2$Fe$_4$O$_9$ and sets the universal unidirectional anisotropy along the negative direction of the applied field. This loss of randomness and preferred orientation of the local anisotropy breaks the inversion symmetry of the jump structure in the hysteresis loop. The net interface moment from this textured anisotropy and development of unidirectional anisotropy of the interface moment toward negative direction of the applied field as a consequence lies at the heart of the path dependency of exchange bias. In fact, this asymmetric jump structure in the hysteresis loop at 2 K is the first clear proof of the presence of textured pattern of anisotropy at the interface between ferromagnetic and antiferromagnetic cores which appears to generate the self generating interface moment along a preferred direction with respect to the direction of the applied field. Topological spin texture in the form of magnetic vortices carrying an electric charge (skyrmion) could earlier be identified in chiral lattice system.\cite{Tokura} The random as well as the textured anisotropy is strong enough as their influence on the hysteresis loop could be seen at a temperature as high as 2 K. In other ferromagnetic systems with local inhomogenity,\cite{Marcano-2} influence of random anisotropy could be seen at even lower temperature (100-500 mK). It is also important to mention here that the influence of this local spin texture in inducing a net interface moment is observable only below the spin freezing temperature  i.e. the Vogel–Fulcher freezing temperature at 29.4K for BiFeO$_3$. The spin structure at the interface needs to be frozen in order to create strong local field. As the temperature is raised toward $\sim$29.4 K from below, influence of the interface spin morphology as well as the net interface moment weaken and so the exchange bias and its path dependency.

The net interface moment from textured anisotropy and consequent path dependency of the exchange bias is quite an attractive proposition as it offers a tunability to the exchange bias depending on the path followed in tracing the hysteresis loop. Since BiFeO$_3$ is a well-known room temperature multiferroic, it is possible to switch the magnetic anisotropy of the ferromagnetic component by applying electric field. Tunable exchange bias then helps in tuning the extent of switching and thus increases the functionality manyfold. 

In summary, we show that a textured pattern of magnetic anisotropy forms at the interface between ferromagnetic Bi$_2$Fe$_4$O$_9$ and antiferromagnetic BiFeO$_3$ nanoparticles from shells of superspin glass and dilute antiferromagnet in a field. The local field from this texture generates net interface moment to set the unidirectional anisotropy along a preferred direction with respect to the direction of the applied field. Such a spontaneous onset of nonswitchable unidirectional anisotropy under field appears to be the origin of the path dependency of the exchange bias. The textured anisotropy at the interface also yields an asymmetric pattern of sharp jumps in the hysteresis loop at low temperature. Thus, a direct correlation could be established between the asymmetric jump structure in the hysteresis loop at low temperature and the path dependency of the exchange bias. Instead of a "clean" interface between ferromagnetic and antiferromagnetic grains, an interface with coexisting local spin texture and random anisotropy emerging out of superspin glass and dilute antiferromagnet in a field, therefore, offers a rare tunability to the exchange bias via its path dependency and, thereby, increases its utility for device applications manyfold.

This work has been supported by ISCA grant no. SFI-12/ISCA/2493, Indo-Ireland joint program of DST, Govt of India (DST/INT/IRE/P-15/11) and Science Foundation of Ireland Principal Investigator Project no. 11/PI/1201. One of the authors (S.G.) acknowledges support in the form of Research Associateship from CSIR, Govt of India, during this work.

\end{document}